\documentclass{PoS}

\newcommand{\en}{\enspace}

\newcommand{\BBfitmassY}{10.90}

\newcommand{\BBfitwidthY}{28 }

\newcommand{\chsqfit}{88/67 }

\newcommand{\wQ}{\it Q}

\title{Tetraquark interpretation of $e^+e^- \to \Upsilon \pi^+\pi^-$
Belle data and $e^+e^- \to b\bar{b}$ BaBar data}

\ShortTitle{Tetraquark interpretation of $Y_b(10890)$ and $R_b$ energy scan}

\author{\speaker{Ahmed Ali}\\
        Deutsches Elektronen-Synchrotron DESY, Notkestrasse 85, D-22607 Hamburg\\
        E-mail: \email{ahmed.ali@desy.de}}

\abstract{We summarize the main features of the spectroscopy, production and decays
of the $J^{PC}=1^{--}$ tetraquarks in the $b\bar{b}$ sector, concentrating on the
lowest state called $Y_b(10890)$. The tetraquark framework is used
to analyze the BaBar data on the $e^+ e^- \to b\bar{b}$ cross section ($R_b$ energy scan) between
$\sqrt{s}= 10.54$ and 11.20 GeV and the Belle data on the processes
$e^+e^- \to \Upsilon(1S) \pi^+\pi^-,\Upsilon(2S) \pi^+\pi^-$ near the peak of the
$\Upsilon(5S)$ resonance. The BaBar $R_b$ energy scan is consistent with an additional state
at a mass of 10.90 GeV and a width of about 28 MeV, in broad agreement with the state $Y_b(10890)$
GeV seen by Belle in the exclusive final states. We argue that the decay widths and
the dipion invariant mass distributions measured by Belle are naturally explained by
the tetraquark interpretation of $Y_b(10890)$.
 }

\FullConference{35th International Conference of High Energy Physics - ICHEP2010,\\
		July 22-28, 2010\\
		Paris France}

\begin{document}
\vspace{-0.3cm}
\section{Introduction}
\vspace{-0.3cm}
Experiments at the $B$ factories and Tevatron have in the past several years 
revived the interest in the spectroscopy of the Quarkonium-like exotic states. 
Labeled tentatively as $X$, $Y$ and $Z$, due to a lack of consensus on their interpretation,
they have masses above the open charm $(D\bar{D})$ threshold,
 with the $X(3872)$ being the lightest and $Y(4660)$ the heaviest 
state observed so far~\cite{Olsen:2009gi}. 
 There is also evidence for an exotic $s\bar{s}$
bound state $Y_s(2175)$ having the quantum numbers $J^{PC}=1^{--}$, first observed in 
the initial state radiation (ISR) process $e^+ e^- \to \gamma_{\rm~ISR}\, f_0(980)\,\phi (1020)$,
where $f_0(980)$ is the $0^{++}$ scalar state. In the $b\bar{b}$ sector, Belle~\cite{Abe:2007tk}
 has observed
enhanced production for the processes $e^+e^- \to \Upsilon(1S) \pi^+\pi^-, \Upsilon(2S)
 \pi^+\pi^-, \Upsilon(3S) \pi^+\pi^-$ in the $e^+e^-$ center-of-mass energy between
$10.83$ GeV and $11.02$ GeV, which does not agree with the conventional $\Upsilon(5S)$
line shape~\cite{Amsler:2008zzb}. The enigmatic features of the Belle data are the anomalously large
decay widths for the mentioned final states and the dipion invariant mass distributions, which
are strikingly different from the conventional QCD expectations for such dipionic transitions.
 A  fit of the Belle data, using a Breit-Wigner resonance, yields a mass  of $10888 ^{+2.7}_{-2.6} ({\rm stat}) \pm 1.2 ({\rm syst})$~MeV and a width of
 $30.7^{+8.3}_{-7.0} ({\rm stat}) \pm 3.1 ({\rm syst})$~MeV~\cite{Abe:2007tk}.
 This particle is given the
tentative name $Y_b(10890)$. In~\cite{Ali:2009es,Ali:2010pq}, $Y_b(10890)$ is interpreted
as a $b\bar{b}$ tetraquark state, which is a linear superposition of the $J^{\rm PC}=1^{--}$
flavour eigenstates $Y_{[bd]}\equiv [bd][\bar{b}\bar{d}]$ and $Y_{[bu]}\equiv [bu][\bar{b}\bar{u}]$. 
The mass eigenstates $Y_{[b,l]}$ (for the lighter) and $Y_{[b,h]}$ (for the heavier)
of the two are almost degenerate, with their small mass difference arising from
 isospin-breaking~\cite{Ali:2009pi}. 
A dynamical model for the decay mechanisms of $Y_b(10890)$ and the final state
distributions measured by
Belle was developed in~\cite{Ali:2009es} and refined in ~\cite{Ali:2010pq}, yielding
good fits of the Belle data. One anticipates that $Y_b(10890)$ is also visible in the
energy scan of the $e ^+ e^- \to b \bar{b}$ cross section, which was undertaken by the BaBar
collaboration between $\sqrt{s}=10.54$ GeV and 11.20 GeV~\cite{:2008hx}.
 A fit of the BaBar data on $R_b$-scan is consistent with a
structure around $Y_b(10890)$ and yields a better $\chi^2/{\rm d.o.f.}$
than the fits without the tetraquark states.  More data are required to resolve this and related
structures in the $R_b$ line shape. 
This contribution summarizes the work done in~\cite{Ali:2009es,Ali:2010pq,Ali:2009pi} interpreting 
the Belle~\cite{Abe:2007tk} and BaBar~\cite{:2008hx} data in terms of the $b\bar{b}$
 tetraquark states.

\vspace{-0.3cm}
\section{Spectrum of bottom diquark-antidiquark states}
\vspace{-0.3cm}
The mass spectrum of tetraquarks $[bq][\overline{bq^{\prime }}]$ with $q=u$, 
$d$, $s$ and $c$ can be described in terms of the constituent diquark
masses, $m_{\wQ}=m_{[bq]}$, spin-spin interactions inside the single diquark, spin-spin
interaction between quark and antiquark belonging to two diquarks,
spin-orbit, and purely orbital term \cite{Drenska:2008gr}, i.e., with a Hamiltonian
\begin{equation}
H=2m_{\wQ}+H_{SS}^{(\wQ\wQ)}+H_{SS}^{(\wQ\bar{\wQ}\mathcal{)}}+H_{SL}+H_{LL} ,
\label{01}
\end{equation}%
where:%
\vspace{-0.3cm}
\begin{eqnarray}
H_{SS}^{(\wQ\wQ)} &=&2(\mathcal{K}_{bq})_{\bar{3}}[(\mathbf{S}_{b}\cdot \mathbf{S%
}_{q})+(\mathbf{S}_{\bar{b}}\cdot \mathbf{S}_{\bar{q}})],  \nonumber \\
H_{SS}^{(\wQ\bar{\wQ}\mathcal{)}} &=&2(\mathcal{K}_{b\bar{q}})(\mathbf{S}%
_{b}\cdot \mathbf{S}_{\bar{q}}+\mathbf{S}_{\bar{b}}\cdot \mathbf{S}_{q})+2%
\mathcal{K}_{b\bar{b}}(\mathbf{S}_{b}\cdot \mathbf{S}_{\bar{b}})+2\mathcal{K}%
_{q\bar{q}}(\mathbf{S}_{q}\cdot \mathbf{S}_{\bar{q}}),  \nonumber \\
H_{SL} &=&2A_{\wQ}(\mathbf{S}_{\wQ}\cdot \mathbf{L}+\mathbf{S}_{%
\bar{\wQ}}\cdot \mathbf{L}),  
\enspace\enspace\enspace\enspace
H_{LL} 
=
B_{\wQ}\frac{L_{\wQ\bar{\wQ}}(L_{\wQ\bar{\wQ}}+1)}{2}.  \label{02}
\vspace{-0.3cm}
\end{eqnarray}%
Here $(\mathcal{K}_{bq})_{\bar{3}}$ is the coupling of the spin-spin interaction between the quarks
inside the diquarks,
$\mathcal{K}_{b\bar{q}}$ are the spin-spin couplings ranging outside the diquark
shells, $A_{\wQ}$ is the spin-orbit coupling of diquark and $B_{\wQ}$
characterizes the contribution of the total angular momentum of the
diquark-antidiquark system to its mass. 

The parameters involved in the above Hamiltonian (\ref{02}) can be obtained
from the known meson and baryon masses by resorting to the constituent quark
model~\cite{De Rujula:1975ge}:
$H=\sum
_{i}m_{i}+\sum
_{i<j}2\mathcal{K}_{ij}(\mathbf{S}_{i}\cdot \mathbf{S}_{j})$,
where the sum runs over the hadron constituents. The coefficient $\mathcal{K}%
_{ij}$ depends on the flavour of the constituents $i$, $j$ and on the particular
colour state of the pair. Using the entries in the PDG~\cite{Amsler:2008zzb} for hadron masses along
with the assumption that the spin-spin interactions are independent of
whether the quarks belong to a meson or a diquark, the results for the
masses corresponding to the tetraquarks $[bq][\bar{b}\bar{q}]$ ($q=u,d,s,c$) were
calculated in \cite{Ali:2009pi}. The lowest eight $1^{--}$
tetraquark states $[bq][\bar{b}\bar{q}]$ ($q=u, d)$, which
are  all orbital excitations with $L_{\wQ \bar{\wQ}}=1$, have the following spin and
orbital angular momentum eigenvalues:
$
Y_{[bq]}^{(1)}\left( S_{\wQ}=0,~S_{\bar{\wQ}}=0,~S_{\wQ\bar{\wQ}}=0,~L_{\wQ\bar{\wQ}}=1\right)$,
$Y_{[bq]}^{(2)}\left( S_{\wQ}=1,~S_{\bar{\wQ}}=0,~S_{\wQ\bar{\wQ}}=1,~L_{\wQ\bar{\wQ}}=1\right)$,
$Y_{[bq]}^{(3)}\left( S_{\wQ}=1,~S_{\bar{\wQ}}=1,~S_{\wQ\bar{\wQ}}=0,~L_{\wQ\bar{\wQ}}=1\right)$,
and\\
$Y_{[bq]}^{(4)}\left( S_{\wQ}=1,~S_{\bar{\wQ}}=1,~S_{\wQ\bar{\wQ}}=2,~L_{\wQ\bar{\wQ}}=1\right)$. 
Identifying the lowest lying $J^{PC}=1^{--}$ state 
$Y_{[bq]}^{(1)}$ with the $Y_b(10890)$ measured by Belle, and using the estimates for the
other parameters entering in Eq.~(\ref{02}), fixes the diquark mass
$m_{\wQ}=m_{\left[ bq\right]}=5.251$ GeV.  The uncertainties on the masses of the other
six states $ M_{Y_{[bq]}}^{(n)}$ ($n=2,3,4 $) are higher, as they depend in addition on
the mass-splittings between the {\it good} and {\it bad} diquarks,
$\Delta=m_{\wQ} (S_{\wQ}=1)- m_{\wQ}(S_{\wQ}=0)$, estimated as
 $\Delta \simeq 200$ MeV~\cite{Jaffe:2003sg,Jaffe:2004ph}. The central values of their masses
are:
$ M_{Y_{[bq]}}^{(2)}=11133$ MeV, $ M_{Y_{[bq]}}^{(3)}=11257$ MeV, and $ M_{Y_{[bq]}}^{(4)}=11227$ MeV.
Assuming isospin symmetry, the states $Y_{[bu]}^{(n)}$ and $Y_{[bd]}^{(n)}$ are degenerate for
each $n$. Including isospin-symmetry breaking lifts this degeneracy with the  mass difference
between the lighter and the heavier of the two states estimated as $M(Y^{(n)}_{[b,l]})
- M(Y^{(n)}_{[b,h]})=(7\pm 3 ) \cos (2 \theta)$ MeV, where $\theta$ is a mixing angle and the
mass eigenstates are defined as: $Y^{(n)}_{[b,l]}=\cos \theta Y^{(n)}_{[bd]} + \sin \theta  Y^{(n)}_{[bu]}$
and $Y^{(n)}_{[b,h]}=-\sin \theta Y^{(n)}_{[bd]} + \cos \theta  Y^{(n)}_{[bu]}$\cite{Ali:2009pi}.
The resulting mass differences are small. However, depending on $\theta$, the
electromagnetic couplings of the $Y^{(n)}_{[b,l]} $ and $Y^{(n)}_{[b,h]}$ may turn out to be
significantly different from each other, and hence also their contributions to $R_b$.
\vspace{-0.3cm}
\section{Decay Widths of $Y_b(10890)$ and other $J^{PC}=1^{--}$ tetraquarks}
\vspace{-0.3cm}
As the masses of all the eight $J^{PC}=1^{--}$ $[bq][\bar{b}\bar{q}]$ tetraquark states lie
 above the thresholds for the decays $Y^{(n)}_{[bq]} \to B_q^{(*)} \bar{B}_q^{(*)}$, they
decay readily into these final states. For the $n=3$ state (having a mass of 11257 MeV),
 also the decay
 $Y^{(3)}_{[bq]}\to \Lambda_b \overline{\Lambda_b}$ is energetically allowed.  In~\cite{Ali:2009pi},
the decay widths $\Gamma(Y^{(n)}_{[bq]} \to B_q^{(*)} \bar{B}_q^{(*)})$ have been estimated
(up to a tetraquark hadronic size parameter $\kappa$) in terms of the corresponding partial
 decay widths  $\Gamma(\Upsilon(5S) \to  B_q^{(*)} \bar{B}_q^{(*)})$, which can be calculated
 with the help of the entries in the PDG~\cite{Amsler:2008zzb}. Specifically, the following relations are assumed
 \vspace{-0.2cm}
\begin{equation}
\kappa^2 \langle B^+B^- |\hat{H}|Y^{(n)}_{[bu]}\rangle= \kappa^2 \langle B^0\bar{B^0}
 |\hat{H}|Y^{(n)}_{[bd]}\rangle= \langle B^+B^- |\hat{H}|\Upsilon(5S)\rangle=
\langle B^0\bar{B^0}|\hat{H}|\Upsilon(5S)\rangle~,
\vspace{-0.2cm}
\end{equation}
and likewise for the $B\bar{B^*}$ and $B^* \bar{B^*}$ decays. Noting that the decays
$\langle B^+B^- |\hat{H}|Y^{(n)}_{[bd]}\rangle$, $\langle B^0\bar{B^0}||\hat{H}|Y^{(n)}_{[bu]}\rangle$
as well as the decays
 $\Gamma(Y^{(n)}_{[bq]} \to B_s^{(*)} \bar{B}_s^{(*)})$ are Zweig-forbidden, one expects,
concentrating on the lowest mass state, 
$\Gamma(Y^{(1)}_{[bq]}) \simeq 0.4 \Gamma(\Upsilon(5S)$. Using the PDG value~\cite{Amsler:2008zzb} 
$\Gamma(\Upsilon(5S)=110$ MeV, we get $\Gamma(Y^{(1)}_{[bd]})=  \Gamma(Y^{(1)}_{[bu]}=
(44 \pm 8)\kappa^2$ MeV for the total decay widths. Equating this decay width to the measured value
 of the total decay width
 $\Gamma[Y_b(10890)]=30.7^{+8.3}_{-7.0} ({\rm stat}) \pm 3.1 ({\rm syst})$ MeV by
Belle~\cite{Abe:2007tk}, one gets
$\kappa= \sqrt{\frac{28 \pm 2}{44 \pm 8}}=0.8 \pm 0.1$. This suggests that the tetraquarks
$Y^{(n)}_{[bq]}$  have a hadronic size of the same order as that of the $\Upsilon(5S)$.
The hadronic widths of the other $J^{PC}=1^{--}$ tetraquarks are estimated as~\cite{Ali:2009pi}:
$\Gamma(Y^{(2)}_{[bq]})= 80 \pm 16$~MeV, $\Gamma(Y^{(3)}_{[bq]})= 114 \pm 22$~MeV and
$\Gamma(Y^{(4)}_{[bq]})= 102 \pm 20$~MeV. 

To calculate the production cross sections, we have derived the
corresponding Van Royen-Weisskopf formula for the leptonic decay
widths of the tetraquark states made up of point-like diquarks~\cite{Ali:2010pq}:
\vspace{-0.3cm}
\begin{eqnarray}
\Gamma( Y_{[bu/bd]} \to e^+e^-)
=
\frac{24\alpha^2|Q_{[bu/bd]}|^2}{m_{Y_b}^4}\,
\kappa^2\left|R^{(1)}_{11}(0)\right|^2,
\label{eq:VRW-P}
\vspace{-0.3cm}
\end{eqnarray}
where $\alpha$ is the fine-structure constant, $Q_{[bu]}=+1/3$, $Q_{[bd]}=-2/3$ are the  diquark charges
 in units of the proton electric charge, 
and $| R^{(1)}_{11}(0)|^2=2.067$ GeV$^5$~\cite{Eichten:1995ch}
is the square of the derivative of the radial wave function for
$\chi_b(1P)$ taken at the origin. Hence, the leptonic widths of the
tetraquark states are estimated as~\cite{Ali:2010pq} 
\vspace{-0.2cm}
\begin{eqnarray}
\Gamma( Y_{[bd]} \to e^+e^-) 
= 4\,\Gamma( Y_{[bu]} \to e^+e^-) 
\approx 83\, \kappa^2\ {\rm eV}\,,
\label{eq:Ybtoee_value}
\vspace{-0.3cm}
\end{eqnarray}
which are substantially smaller than the leptonic width of the
$\Upsilon(5S)$~\cite{Amsler:2008zzb}. The electronic widths of the mass eigenstates
$Y_{[b,l]}$ and $Y_{[b,h]}$ depend, in addition, on the mixing angle $\theta$.
\vspace{-0.3cm}
\section{Analysis of the BaBar data on $R_b$-scan}
\begin{figure}[t]
\begin{center}
\includegraphics[width=0.45\textwidth]{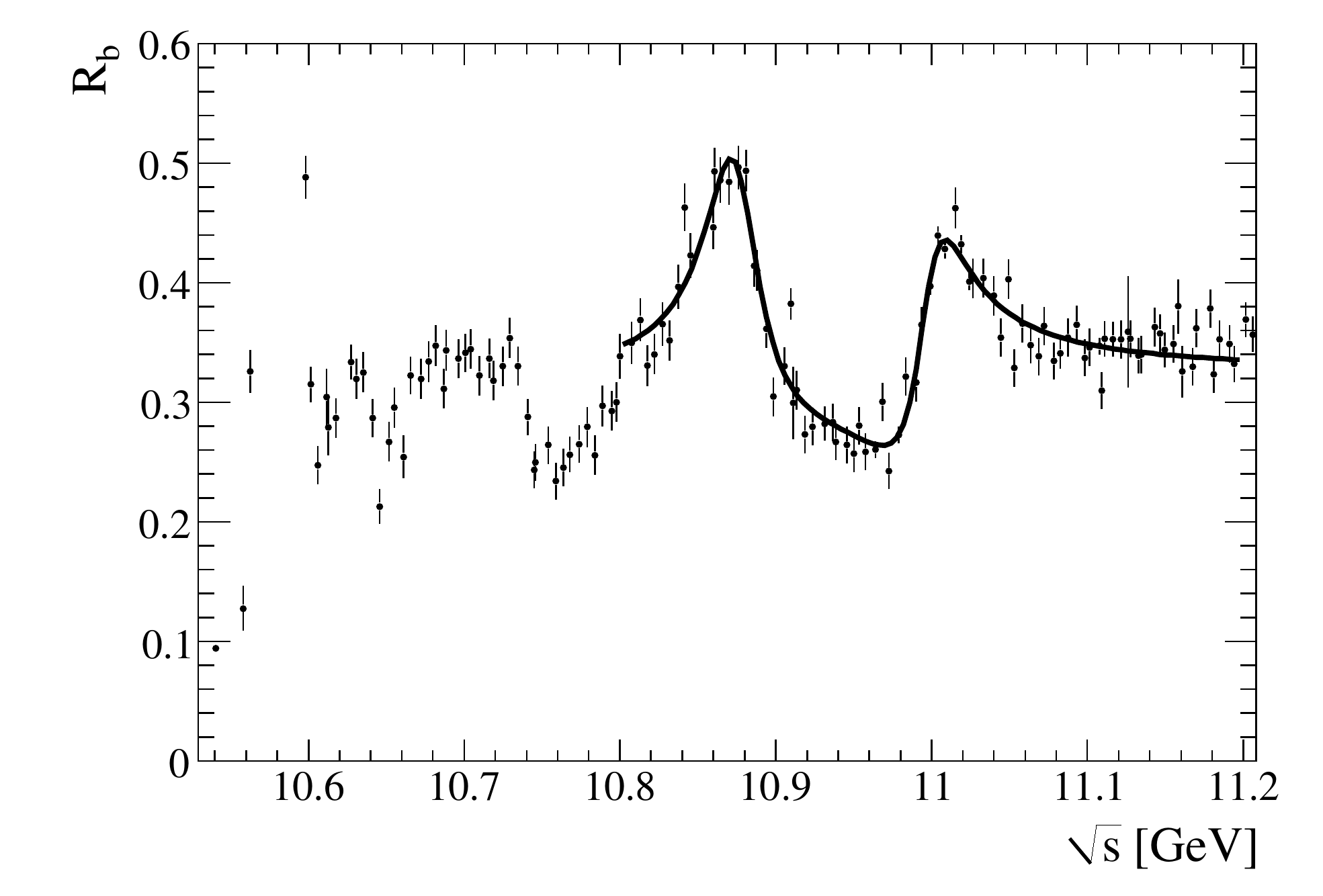}
\includegraphics[width=0.425\textwidth]{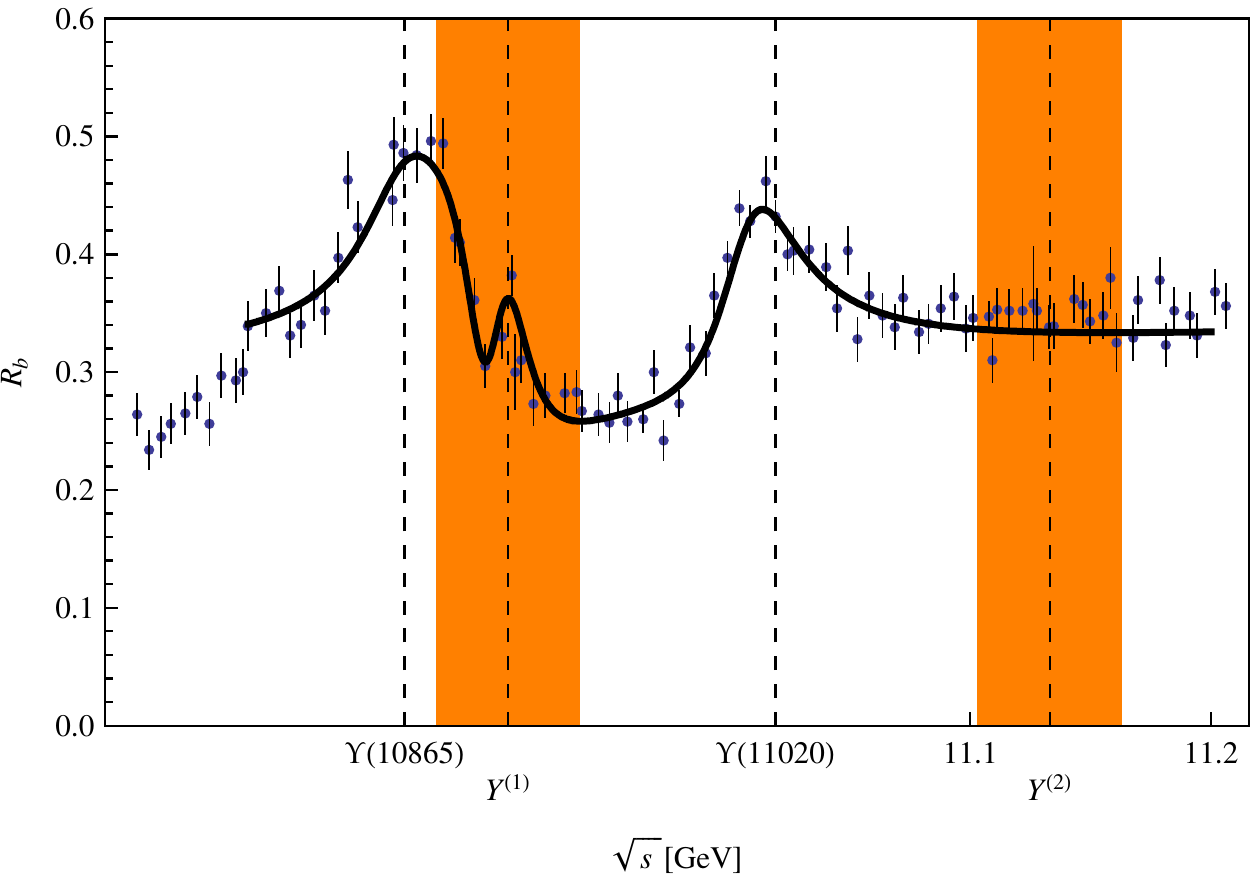}
\vspace*{-3mm}
\caption{\label{BABARPlots}
Measured $R_b$ as a function of $\sqrt{s}$ with the result of the fit with 2 Breit-Wigners described in
Fig.~1 of B.~Aubert {\it et al.}  [BaBar Collaboration] \cite{:2008hx} (left-hand frame). 
The result of the fit of the $R_b$ data with 4 Breit-Wigners~\cite{Ali:2009pi} (right-hand frame).
Location of the $\Upsilon (5S)$,  $\Upsilon (6S)$, the tetraquark states
$Y_{[b,q]}^{(1)}$ (labelled as $Y^{(1)}$) and 
$Y_{[b,q]}^{(2)}$ (labelled as $Y^{(2)}$) are indicated. The shaded bands around
the mass of $Y^{(1)}$ and $Y^{(2)}$ reflect the theoretical uncertainty in their masses.
(From ~\cite{Ali:2009pi}).}
\end{center}
\end{figure}

\vspace{-0.3cm}
BaBar has  reported the $e^+ e^- \to b \bar{b}$ cross section measured in a dedicated 
 energy scan  in the range $10.54$~GeV and $11.20$~GeV taken in steps of 5~MeV~\cite{:2008hx}.
Their measurements are shown in  Fig.~\ref{BABARPlots} (left-hand frame) together with the result of the
 BaBar fit which contains the following ingredients:
A flat component representing the $b\bar{b}$-continuum states not interfering with resonant decays,
called $A_{nr}$, added incoherently to a second flat component, called $A_r$, interfering with two relativistic
Breit-Wigner resonances, having the amplitudes  $A_{10860}$, $A_{11020}$ 
and strong phases, $\phi_{10860}$ and  $\phi_{11020}$, respectively. Thus,
\vspace{-0.3cm}
\begin{eqnarray}
\label{babarfit-simple}
\sigma (e^+e^-\to b \bar{b}) &=& \vert A_{nr}\vert^2 + \vert A_r + A_{10860}e^{i \phi_{10860} } BW(M_{10860}, \Gamma_{10860})
\nonumber
\\
&&
+ A_{11020}e^{i \phi_{11020} } BW(M_{11020}, \Gamma_{11020})\vert^2~,
\vspace{-0.4cm}
\end{eqnarray}
with $BW(M,\Gamma)=1/[(s-M^2) +i M \Gamma]$. The results summarized in their Table II for the masses
and widths of the $\Upsilon(5S)$ and $\Upsilon(6S)$ differ substantially from the corresponding
PDG values~\cite{Amsler:2008zzb}, in particular, for the widths, which are found to be
$43 \pm 4$~MeV for the $\Upsilon(10860)$, as against the PDG value of $110 \pm 13$~MeV,
and $37\pm 2$~MeV for the $\Upsilon(11020)$, as compared to $79\pm 16$~MeV in PDG.
 As the systematic errors  from the various thresholds are not taken into account,
this mismatch needs further study. 
The fit shown in Fig.~\ref{BABARPlots} (left-hand frame) is not particularly impressive
having a $\chi^2/{\rm d.o.f.}$ of approximately~2. 

The BaBar $R_b$-data is refitted in~\cite{Ali:2009pi} by modifying the model in
 Eq.~(\ref{babarfit-simple}) by taking into account
two additional resonances, corresponding to the masses and widths of $Y_{[b,l]}$ and $Y_{[b,h]}$. Thus,
 formula (\ref{babarfit-simple}) is extended by two more terms 
 \vspace{-0.2cm}
\begin{equation}
A_{Y_{[b,l]}}e^{i \phi_{Y_{[b,l]}} } BW(M_{Y_{[b,l]}}, \Gamma_{Y_{[b,l]}})
\en\en\en\textnormal{     and     } \en\en\en
A_{Y_{[b,h]}}e^{i \phi_{Y_{[b,h]}} } BW(M_{Y_{[b,h]}}, \Gamma_{Y_{[b,h]}}),
\vspace{-0.2cm}
\end{equation}
which interfere with the resonant amplitude $A_{r}$
  and the two resonant amplitudes for $\Upsilon (5S)$ and $\Upsilon (6S)$
 shown in 
Eq.~(\ref{babarfit-simple}). Using the same non-resonant amplitude $A_{nr}$ and $A_r$ 
as in the BaBar analysis~\cite{:2008hx}.
the resulting fit is shown in Fig.~\ref{BABARPlots} (right-hand frame).
Values of the best-fit parameters yield 
 the masses of the $\Upsilon(5S)$ and $\Upsilon(6S)$ and their respective full widths 
 which are almost identical to the values obtained by BaBar~\cite{:2008hx}. 
 However, quite strikingly,  a third resonance is seen in the
 $R_b$-line-shape at a mass of $\BBfitmassY$~GeV,
 tantalisingly close to the $Y_b(10890)$-mass in the Belle measurement
of the cross section for $e^+e^- \to Y_b(10890) \to \Upsilon(1S,2S)\; \pi^+\pi^-$, and a width 
of about $\BBfitwidthY$~MeV. 
In the region around 11.15 GeV, where the $Y^{(2)}_{[bq]}$ states are expected, our fits
 of the BaBar $R_b$-scan do not show a resonant structure due to the larger decay widths
 of the states
$Y^{(2)}_{[bq]}$. The resulting $\chi^2/{\rm d.o.f.} = \chsqfit$ with the 4 Breit-Wigners shown 
in Fig.~\ref{BABARPlots} (right frame) is  better
than that of the BaBar fit~\cite{:2008hx}.
 
 The quantity $\mathcal{R}_{ee}(Y_b)= \Gamma_{ee}(Y_{[b,l]})/\Gamma_{ee}(Y_{[b,h]})$  is
 given by the ratio of the two amplitudes
 $A_{Y_{[b,l]}}$ and $A_{Y_{[b,h]}}$, which also fixes the mixing angle $\theta$. From the fit
shown in the right-hand frame in Fig.~\ref{BABARPlots},  one obtains  
 \vspace{-0.2cm}
 \begin{equation}
 \mathcal{R}_{ee}(Y_b)=1.07 \pm 0.05,
 \vspace{-0.2cm}
 \end{equation}
yielding  
\vspace{-0.2cm}
\begin{equation}
\theta=-19 \pm 1 ^{\circ}\en\en\en\textnormal{and}\en\en\en \Delta M=5.6 \pm 2.8\;\textnormal{MeV},
\vspace{-0.2cm}
\end{equation}
for the  mixing angle and  the mass difference between the eigenstates, respectively.
For the mass eigenstates $Y_{[b,l]}$ and  $Y_{[b,h]}$, the electronic widths $\Gamma_{ee}(Y_{[b,l]})$
and $\Gamma_{ee}(Y_{[b,h]})$ are given by~\cite{Ali:2010pq}
$\Gamma_{ee}(\theta)=0.2\; \kappa^2 Q(\theta)^2$~keV. With the above determination of
$\kappa$ and $\theta$, we get  
\begin{equation}
\Gamma_{ee}(Y_{[b,l]})=0.033\pm 0.006 \; \textnormal{keV}\en\en\en\textnormal{and}\en\en\en \Gamma_{ee}(Y_{[b,h]})=0.031\pm 0.006\; \textnormal{keV}.
\end{equation}
\vspace{-3mm}
\section{Analysis of the Belle data on $e^+e^- \to Y_b \to (\Upsilon(1S),\Upsilon(2S)) \pi^+\pi^-$}
\begin{figure}[t]
\centering
\resizebox{0.95\textwidth}{0.25\textwidth}{
\includegraphics[width=0.8\textwidth,height=10cm]{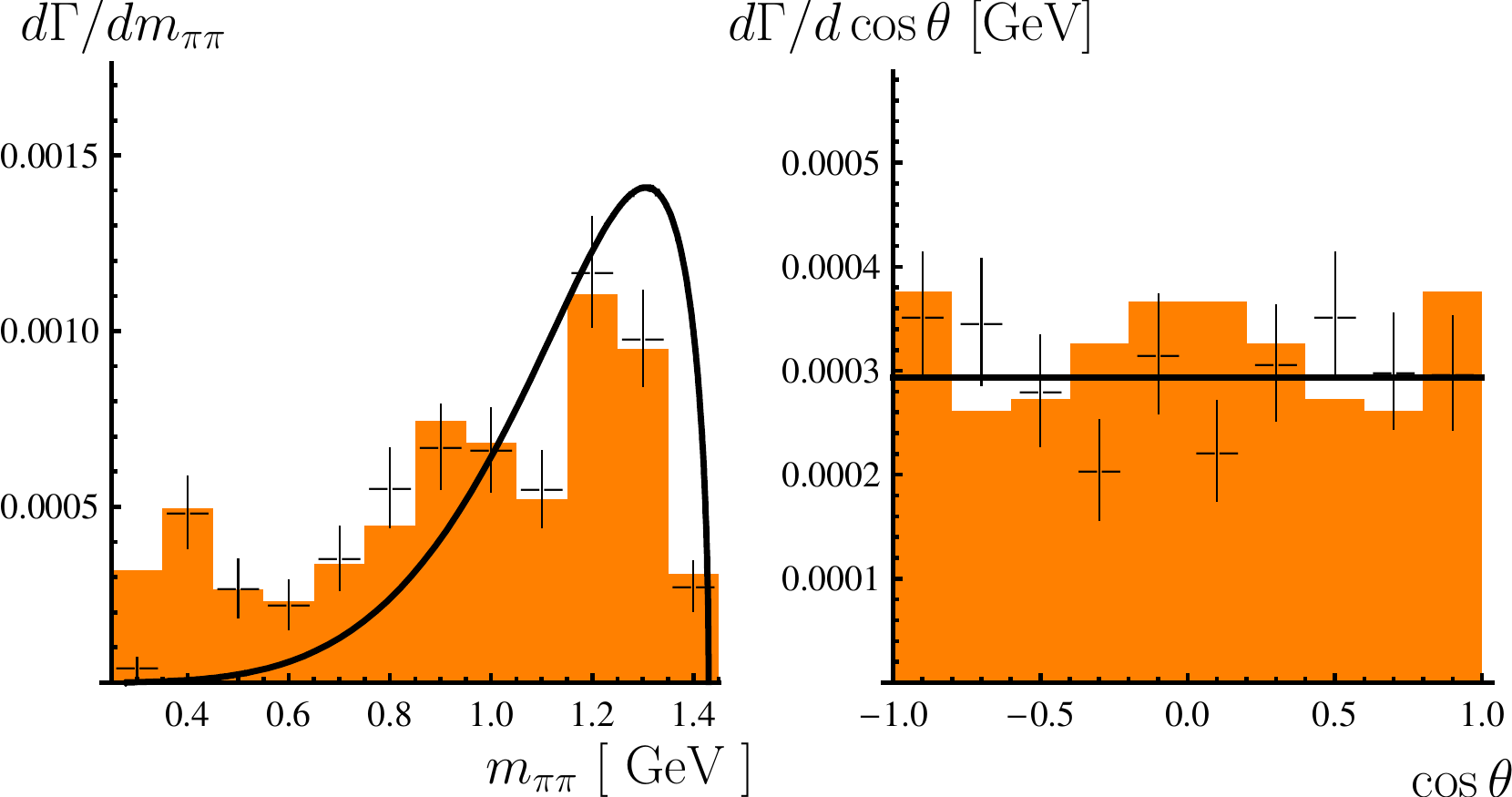}
\includegraphics[width=0.8\textwidth,height=10cm]{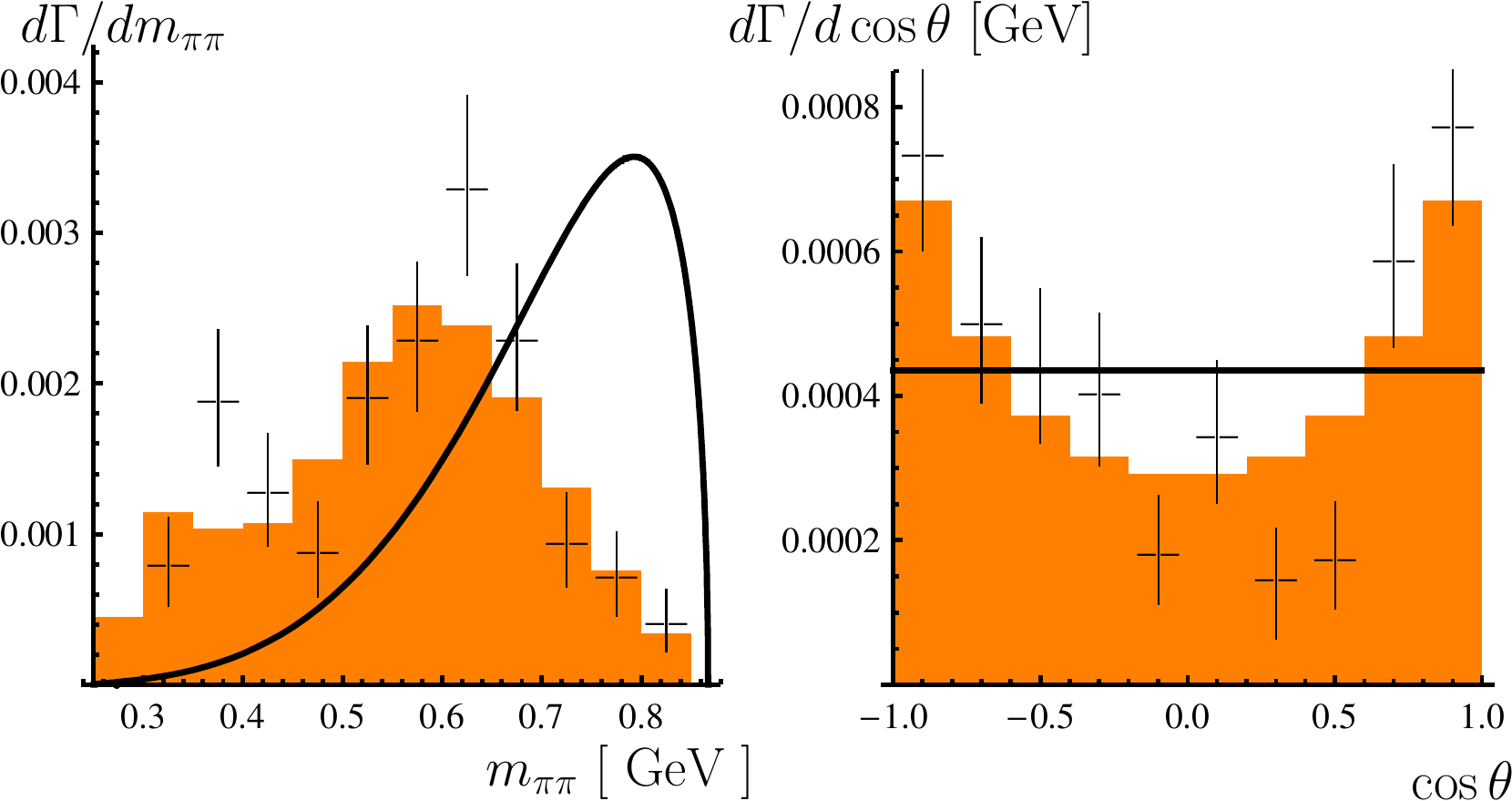}
}
\vspace*{-3mm}
\caption{Left-hand frames: Fit results of the $M_{\pi^+\pi^-}$ distribution  and  the
  $\cos\theta$ distribution for $e^+ e^- \to Y_b \to \Upsilon(1S) \pi^+\pi^-$, 
 normalized by the measured cross section by  Belle~\cite{Abe:2007tk}. Right-hand frames:
 The same distributions for  $e^+e^- \to \Upsilon(2S) \pi^+\pi^-$.
In all figures, the histograms 
  represent the fit results based on tetraquarks, while the crosses are the 
  Belle data~\cite{Abe:2007tk}. 
The solid curves in the figures show purely continuum contributions.
 (From ~\cite{Ali:2009es}.) 
\label{fig:spectra}
}
\end{figure}

\vspace{-3mm}
With the $J^{\rm PC}=1^{--}$ for both $Y_b$ and $\Upsilon(nS)$, the dipionic final state is allowed
 to have the quantum numbers $0^{++}$ and $2^{++}$.  There are
only three low-lying resonances in the PDG which can contribute as intermediate states, namely, the two $0^{++}$ states, $f_0(600)$ and $f_0(980)$, which we take as the lowest tetraquark states, and the
$2^{++}$ $q\bar{q}$ meson state $f_2(1270)$. All
three states contribute for the final state $\Upsilon(1S) \pi^+\pi^-$. However, kinematics allows only the $f_0(600)$ in the final state
$\Upsilon(2S) \pi^+\pi^-$. In addition, a non-resonant contribution with a significant $D$-wave 
fraction is needed by the data on these final states. This model accounts well the shape of the
measured distributions, as shown in Fig.~\ref{fig:spectra} for $e^+e^- \to Y_b \to \Upsilon(1S)
 \pi^+\pi^-$ (left-hand frames) and  for
 $e^+e^- \to Y_b \to \Upsilon(2S) \pi^+\pi^-$ (right-hand frames).
As the decays $Y_b \to (\Upsilon(1S),\Upsilon(2S)) \pi^+\pi^-$ are Zweig-allowed, one expects
larger decay widths for these transitions, typically of $O(1)$
 MeV~\cite{Ali:2010pq}, 
 than the decay widths for the conventional dipionic transitions, such as
$\Upsilon(4S) \to \Upsilon(1S) \pi^+\pi^-$, which are of order 1 keV~\cite{Amsler:2008zzb}.
  Further tests of the tetraquark
hypothesis involving the processes $e^+e^- \to Y_b \to \Upsilon(1S) (K^+ K^-, \eta \pi^0)$
are presented in ~\cite{Ali:2010pq}.

\end{document}